\documentclass{article}
\usepackage{graphicx} 
\usepackage{amsmath}

\usepackage{tikz}
\usetikzlibrary{shapes.geometric, arrows}
\usepackage{hyperref}

\tikzstyle{Dom} = [rectangle, minimum width=1cm, minimum height=1cm, text centered, draw=black, fill=orange!30]
\tikzstyle{Sub} = [rectangle, minimum width=0.6cm, minimum height=0.6cm, text centered, draw=black, fill=orange!30]

\tikzstyle{arrow} = [thick,<-,>=stealth]
\tikzstyle{sarrow} = [thick,-,>=stealth]

\usepackage[
backend=biber,
style=alphabetic,
sorting=ynt
]{biblatex}
\title{POEM: Proof of Entropy Minima}
\author{Karl Kreder \\ \href{mailto:karl@dominantstrategies.io}{karl@dominantstrategies.io} 
   \and Shreekara Shastry \\ \href{mailto:shreekara@dominantstrategies.io} {shreekara@dominantstrategies.io} }

\date{March 7, 2023}

\begin{document}

\maketitle

\begin{abstract}
       Nakamoto consensus has been incredibly influential in enabling robust blockchain systems, and one of its components is the so-called heaviest chain rule (HCR).  Within this rule, the calculation of the weight of the chain tip is performed by adding the difficulty threshold value to the previous total difficulty. Current difficulty based weighting systems do not take the intrinsic block weight into account.  This paper proposes a new mechanism based on entropy differences, named proof of entropy minima (POEM), which incorporates the intrinsic block weight in a manner that significantly reduces the orphan rate of the blockchain while simultaneously accelerating finalization. Finally, POEM helps to understand blockchain as a static time-independent sequence of committed events.
\end{abstract}

\section{Introduction}

    In their seminal work, Nakamoto \cite{Nakamoto2008} introduces a novel form of consensus which is often referred to as "Nakamoto Consensus". Among its core attributes,   the choice of the canonical tip of a blockchain is made by selecting the "heaviest" head based on a particular implementation of the so-called heaviest chain rule (HCR).  This arguably led to the first economic byzantine fault tolerant \cite{Lamport1982} mechanism for coordinating an open group of node operators in a distributed system. 

    However, Nakamoto consensus using this implementation of HCR suffers from certain disadvantages, one of which is the production of orphaned blocks resulting from  propagation delays (and thus partial information) within the network.  Orphaned blocks are valid blocks that share a common parent block but have two different commitments with two different but equally valid proofs. However, only one of the blocks can ultimately be accepted by the network as canonical.  As the network participants do not know which block(s) were produced first, they  assume an ordering based on the order of the receipt of each block.  Since both blocks are assigned the threshold difficulty weight and share a common parent, there is no alternative preferred objective mechanism for picking the head. Therefore, the network cannot converge on the choice of canonical head until one or more additional blocks are found. This can lead to delays in consensus each time such an event occurs. This problem becomes especially pernicious within blockchains with consistently high orphan block rates, as orphans lead to the production of more orphans. Moreover, this results in discarded hashes generated by the associated Proof-of-Work (PoW) algorithm used within Nakamoto Consensus. An attempt at addressing wasted work in blockchain production with many orphans was proposed by \cite{Sompolinsky2015} with the greediest heaviest observed sub-tree (GHOST). GHOST includes a discounted addition of weight for orphaned blocks, referred to as uncles within GHOST, that are referenced by the head.  Inclusion of uncles in the weight calculation helps to better measure the total work referenced by the various head choices in a noisy environment.
    
  PoW based blockchains achieve economic finalization at the point when the cost of an attack exceeds the benefit to the attacker. The practical exploitation of finalization latency is manifested by attackers mining private chains which they eventually reveal to revert one or more transactions. This is known as a 51\% attack but \cite{Eyal2013} shows that this can be reduced to 33\% with coordination of mining pools. In addition to 51\% attacks, nefarious miners can cause the reversion of large amounts of work through block-withholding attacks \cite{Courtois2014}.  

    As throughput is one of the primary limitations faced by current blockchain systems, many proposals have been made on scaling blockchains via sharding. A subset of these proposals is applicable to work based consensus mechanism including BlockReduce \cite{Georghiades2022}, treechains, fruitchains \cite{Pass2017}, and bitcoinNG \cite{Metropolitana1940}. Each of these solutions propose the use of sub-chains or shares as a mechanism to asynchronously produce datasets prior to inclusion in the highest level block data structure.  However, subchains or shares make these proposals more vulnerable to withholding attacks due to block weight stratification for the various block types in the hierarchy. The Hierarchical Longest Chain Rule (HLCR)\cite{Georghiades2022}, an enhancement on HCR described above, proposes that a tip is chosen hierarchically such that subchain choices are restricted by the longest tip in a dominant chain.  Although promising, HLCR is potentially vulnerable to withholding attacks which would cause all of the subchain blocks to be discarded up to the average block time of the most dominant chain. Fruitchains proposes mitigation of withholding attacks by weighting shares based on age at inclusion. This can potentially partially mitigate withholding attacks. Here, we present a potentially more powerful manner of addressing such attacks using an entropic mechanism for consensus, and minimization of difference entropy as a mechanism for head choice. 

\section{Entropic Consensus}

Any blockchain's evolution is that of a random process. In the classical proof-of-work (PoW) setting, an associated PoW function (often a hash function) has $2^l$ states and is (approximately, if well-designed) uniformly distributed across these states. Thus, the maximum entropy associated with hashed outputs is $S_{max} = l$ bits, corresponding to no "work" performed on the system \cite{Shannon1948}.

The PoW algorithm restricts acceptable hashed outputs to all values below a threshold difficulty $2^d$, implying that the first $l-d$ bits of the output hash must be zero. This sets the maximum output target entropy to be $d$ bits.

In practice, the mining process achieves a hashed output that is less than or equal to the difficulty threshold, i.e., it may possess {\em greater} than or equal to $l-d$ leading zeros. We call this the intrinsic difficulty $d_{int}$, resulting in a sequence with $c \le d$ non-zero elements. Thus, in practice, the realized entropy of the output is $c$ bits and the reduction in the entropy is $l-c$ bits which represent the number of leading zeros that will be called $n$. Additionally, $n$ can include fractional zero bits which are after the first non-zero bit. More precisely, this makes $n = l - log_2(d_{int}) $. The intrinsic difficulty can be used to calculate the {\em difference entropy}:
\begin{equation}
\boxed{\Delta S =  \frac{1}{2^n}} \label{eq:1}
\end{equation}
where $\Delta S $ represents the number of possible states removed from the macrostate in the achievement of a single block. This can be extended to compute the change in entropy in an arbitrary sequence of $k$ blocks.

\begin{flalign*}
   &\Delta S_{k} = \Delta S_{k-1} \times \frac{1}{2^{n_k}} \\
   log_2 \Delta S_k & = log_2 \Delta S_{k-1} + log_2(\frac{1}{2^{n_k}}) \\
   & = log_2 \Delta S_{k-1} - n_k     
\end{flalign*}
This allows the computation of the $\Delta S_k $ to simply be carried out by the summation of all prior zero bits found in a chain.
\begin{equation}
   \boxed{log_2 \Delta S_k =  -\Sigma_{i=1}^{k} n_i} \label{eq:2}
\end{equation}

\subsection{Impact on Finalization}
    To understand the impact of using $\Delta S_k$ to determine a canonical blockchain tip, it is most illustrative to consider a system with at least one subchain. For example, consider a dominant blockchain which merge mines a subchain.\footnote{The definitions and a deeper understanding of dominant chain, subchain and merged mining can be found in \cite{Georghiades2022}}

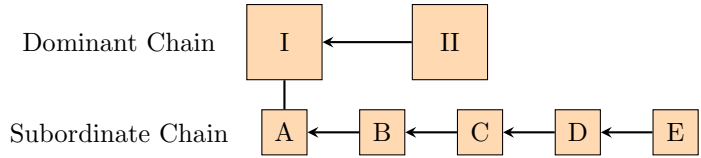
\begin{figure}[h]
\centering
\begin{tikzpicture}[node distance=1.2cm]
\node (d0) [Dom] {I};
\node (d1) [Dom, right of=d0, xshift=1cm] {II};
\node [left of=d0, xshift=-1cm] {Dominant Chain};

\node (s0) [Sub, below of=d0] {A};
\node (s1) [Sub, right of=s0, xshift=0.1cm] {B};
\node (s2) [Sub, right of=s1, xshift=0.1cm] {C};
\node (s3) [Sub, right of=s2, xshift=0.1cm] {D};
\node (s4) [Sub, right of=s3, xshift=0.1cm] {E};
\node [left of=s0, xshift=-1cm] {Subordinate Chain};

\draw [arrow] (d0) -- (d1);
\draw [arrow] (s0) -- (s1);
\draw [arrow] (s1) -- (s2);
\draw [arrow] (s2) -- (s3);
\draw [arrow] (s3) -- (s4);

\draw [sarrow] (d0) -- (s0);

\end{tikzpicture}

\caption{Dominant and Subordinate Chains}
\label{fig:1}
\end{figure}

Let the subordinate threshold difficulty be $m_t$ and the dominant block difficulty be $m_t+m_d$. In a traditional calculation of difficulty based on target difficulty, the dominant blocks that meet (or exceed the target) will be assigned the weight:
\begin{equation*}
    \propto 2^{m_t+m_d} 
\end{equation*} Each of the subordinate blocks would have a difficulty $\propto 2^{m_t}$. Therefore the subordinate chain would be chosen when:

\begin{equation*}
    k \times 2^{m_t} > 2^{m_t + m_d} 
\end{equation*}
\begin{equation}
    \boxed{k > 2^{m_d}} \label{eq:3}
\end{equation}
Alternatively, using the  $\Delta S_k$ formulation, the dominant block's entropy is:
\begin{equation*}
    \Delta S_{dom} = \frac{1}{2^{(m_t+m_d)}}
\end{equation*}
The entropy of the $k^{th}$ subordinate blocks is:
\begin{equation*}
    \Delta S_k = \frac{1}{2^{k(m_t+1)}}
\end{equation*}
Therefore, the subordinate chain would be chosen when:
\begin{equation*}
    k\times m_t > (m_t+m_d)
\end{equation*} 
\begin{equation}
    \boxed{k>\frac{(m_t+m_d)}{m_t}} \label{eq:4}
\end{equation}
The entropy measurement allows the subordinate to overtake the dominant with a linear number of blocks whereas the difficulty measurement would require the subordinate to have an exponential number of blocks.

Practically speaking, if we take $m_t$ = 20 and $m_d$ = 5 the difficulty measurement would require 32 subordinate blocks whereas the entropy measurement would only require 2 blocks. Using the difficulty measurement, a dominant block could be withheld for 32 subordinate blocks or the average dominant block time.  Alternatively, using an entropy measurement, a dominant block could only be withheld for approximately 2 subordinate blocks or 3\% of the dominant block time. This represents a dramatic improvement in the tolerance of merged mined or sub-share blockchain constructions to withholding attacks.  Additionally, this practically allows the sub blocks to accumulate meaningful finalization guarantees prior to inclusion in the dominant chain.

Additionally, there is an interesting impact on the preference in the choice of hash algorithm as well as field size shown by equation \eqref{eq:4}. Although it may be intuitive in the context of POEM, equation \eqref{eq:4} shows that a hash algorithm that is most efficient at reducing entropy while maintaining a collision resistant one-way function with a uniform field distribution is most desirable.  Specifically, given a fixed field size, a higher $m_t$ threshold yields a lower $k$ value and faster finalization guarantees.

\subsection{Actual Difficulty Impact on Orphaned Blockchain}  

POEM enables the removal of the equating the intrinsic block difficulty to the threshold value. Using the intrinsic difficulty will  effectively eliminate competing blocks, and therefore increase the chains hash efficiency.  \\

Let $p(d_{int})$ be the probability of getting a block with the value $d_{int}$. The probability of getting a competing block with the same value of $d_{int}$ is 
\begin{equation}
    p_o(d_{int}) = p(d_{int}) \times \frac{1}{2^l} \label{eq:5}
\end{equation}

Therefore, if two blocks are originated in close temporal proximity, $\frac{2^l - 1}{2^l} $ percent of the time intrinsic difficulty causes one of the blocks to be preferred over the other. Given a sufficiently large choice of $l$ intrinsic difficulty allows the instantaneous resolution of effectively all latency driven forks.

\subsection{Finite Finalization Guarantee}
In the difficulty measurement paradigm, the actual difficulty cannot be used because it has an exponential relative weight compared to the head calculation. When $c-d$ additional bits are found past the threshold, equation \eqref{eq:3} becomes:
\begin{equation*}
    k > 2^{m_d+(c-d)}
\end{equation*}
This means k, although finite, is effectively unbounded and there is no longer any practical finalization guarantee. However, when the entropy calculation is used, the intrinsic difficulty can also be used as they are both of linear weight in a logarithmic field. When $c-d$ additional bits are found past the threshold, equation \eqref{eq:4} becomes:
\begin{equation*}
    k > \frac{m_t+m_d+(c-d)}{m_t}
\end{equation*}
Moreover, when coupled with the entropy calculation, the finalization duration remains finite and is bounded by the bits in the hash function field. For a 256-bit field this would guarantee $1 < k < 256$. However, practical choices of $m_t$ would keep $k$ in low single digits.

\section{Discussion}
Independently, the concept of optimizing (difference) entropy as a means of progressing a blockchain is very coherent and almost obvious once presented.  However, it is hard to develop an intuitive understanding of the proposal when bringing certain precepts which are fundamentally incompatible.  Specifically, a blockchain is primarily a sequencer and is completely independent of time. Therefore, any introductions of assumptions which reincorporate time will create a conflicting intuition about the entropy based concept.  Specifically, any considerations of averages, distributions, hash-rate, or time will cause a misunderstanding of this proposal. Ultimately, understanding the nuance of why time based considerations seemingly conflict with the entropy based system is the key to understanding both.  

The entropy calculation provides a new intuition on the relative value of shares, dominant blocks, and interlinks in a blockchain.  Previous intuition, using a work based calculation, would cause the weight of a given block to be proportional to the average hash rate needed to find that single block.  However, this is just a single sample of a large number of independent events. With one sample, the weight assigned to any one block effectively has infinite variance.  This imprecision in measurement and over assignment of single block weight is what causes almost all consensus inefficiencies and attack vectors that exist in PoW blockchains.  Within the entropy context, a dominant block which has 5 additional bits of entropy and is 32 times harder to find than a sub block with 20-bits would only be counted as having an entropy that is 25\% greater.  This makes sense when you consider the chain to be a sequence of references to each other and the tip being the choice that statistically has the greatest amount of measured effort. Only when the dominant block is included in the chain, and also has the expected number of sub shares which it references, does the dominant block get the full bit consideration. Effectively the entropy method takes in all information and creates many sample calculations on the effort applied to any given tip. Thus, POEM makes a choice that most closely reflects reality with the least amount of variance. Therefore, it is likely POEM can solve a number of outstanding problems with current blockchains including selfish mining, withholding attacks, and Sybil resistance $<51\%$ while also allowing sub-share based systems to operate robustly and efficiently. 

Another interesting consequence of using the entropy-based measurement is that unlike difficulty-based systems with subshares the tip of the chain is dictated by the most recent subshare rather than the most dominant block.  This means that instead of the tip being coerced from the top down \cite{Georghiades2022}, the dominant chains are pulled along by the subshares and eventually come into agreement.  Therefore, blocks don't really have high independent $\Delta S$ on their own but rather become canonical as the amount of entropy reduction which references them continues to increase.  This concept of eventually finalizing a block with the increase in depth of a blockchain now holds true even for blockchains with subshares and interlinks.

It appears to the authors that all prior worked based blockchain proposals misrepresented and misunderstood the mechanism being used to choose and extend the tip of a chain. Although this work is closely related with PoW consensus, POEM is a subset within work mechanisms.  Not all work functions are explicitly compatible with POEM and certain functions and field sizes will potentially have preference over others.  Additionally, the lack of precision in the measurement of work as a proxy for entropy has led to many engineering approximations to compensate for the deficiency.  This includes limiting reorgs to a certain depth, coordinating heads by time, and truncating bits at threshold values. With the new precision of using entropy measurements, we believe that all further "choices" in blockchain design may become rationally emergent.

\section{Conclusion}
In this work, a novel consensus mechanism is proposed Proof-of-entropy minima (POEM). It was shown that POEM is able to decrease the order of the time to finalization in a blockchain with sub-shares from exponential to linear.  Additionally, it was shown that POEM can immediately resolve practically all contentious blocks at the blockchain tip. POEM gives closed-form equations to blockchain systems that can be used to analyze the difference between blockchain systems, selfish mining attacks, etc. It is posited that POEM will also be able to address deficiencies in PoW designs which allow for selfish mining and reduction of Sybil resistance below 51\%, however, the specific analysis is left for future work.

\printbibliography

@article{Eyal2013,
   author = {Ittay Eyal and Emin Gun Sirer},
   month = {11},
   title = {Majority is not Enough: Bitcoin Mining is Vulnerable},
   year = {2013},
}

@article{Pass2017,
   author = {Rafael Pass and Elaine Shi},
   doi = {10.1145/3087801.3087809},
   isbn = {9550151026},
   journal = {Proceedings of the ACM Symposium on Principles of Distributed Computing - PODC '17},
   pages = {315-324},
   pmid = {19644474},
   title = {FruitChains: A Fair Blockchain},
   volume = {2016},
   url = {https://doi.org/10.1145/3087801.3087809},
   year = {2017},
}

@inproceedings{Metropolitana1940,
   author = {Zona Metropolitana and Nuevo Le and Baja California and Ciudad Ju and Corredor Industrial Ta},
   doi = {abs/1510.02037},
   isbn = {978-1-931971-29-4},
   journal = {Proceedings of the 13th USENIX Symposium on Networked Systems Design and Implementation (NSDI ’16)},
   title = {Bitcoin-NG: A Scalable Blockchain Protocol},
   year = {1940},
}

@inproceedings{Georghiades2022,
   author = {Yanni Georghiades and Karl Kreder and Jonathan Downing and Alan Orwick and Sriram Vishwanath},
   doi = {10.1109/Blockchain55522.2022.00072},
   isbn = {978-1-6654-6104-7},
   journal = {2022 IEEE International Conference on Blockchain (Blockchain)},
   month = {8},
   pages = {468-475},
   publisher = {IEEE},
   title = {Scalable Multi-Chain Coordination via the Hierarchical Longest Chain Rule},
   year = {2022},
}

@article{Nakamoto2008,
   author = {Satoshi Nakamoto},
   doi = {10.1007/s10838-008-9062-0},
   isbn = {978-972-757-716-3},
   issn = {09254560},
   issue = {1},
   journal = {Journal for General Philosophy of Science},
   pages = {53-67},
   pmid = {14533183},
   title = {Bitcoin: A Peer-to-Peer Electronic Cash SyNakamoto, S. (2008). Bitcoin: A Peer-to-Peer Electronic Cash System. Consulted, 1–9. doi:10.1007/s10838-008-9062-0stem},
   volume = {39},
   year = {2008},
}

@article{Courtois2014,
   author = {Nicolas T. Courtois and Lear Bahack},
   month = {1},
   title = {On Subversive Miner Strategies and Block Withholding Attack in Bitcoin Digital Currency},
   year = {2014},
}

@article{Lamport1982,
   author = {Leslie Lamport and Robert Shostak and Marshall Pease},
   issue = {3},
   journal = {ACM Transactions on Programming Languages and System},
   pages = {382-401},
   title = {The Byzantine Generals Problem},
   volume = {4},
   year = {1982},
}

@article{Sompolinsky2015,
   author = {Yonatan Sompolinsky and Aviv Zohar},
   doi = {10.1007/978-3-662-47854-7_32},
   pages = {507-527},
   title = {Secure High-Rate Transaction Processing in Bitcoin},
   year = {2015},
}

@article{Shannon1948,
   author = {C. E. Shannon},
   doi = {10.1002/j.1538-7305.1948.tb01338.x},
   issn = {00058580},
   issue = {3},
   journal = {Bell System Technical Journal},
   month = {7},
   pages = {379-423},
   title = {A Mathematical Theory of Communication},
   volume = {27},
   year = {1948},
}

\end{document}